\documentclass{llncs}

\AtBeginDocument{%
  \providecommand\BibTeX{{%
    \normalfont B\kern-0.5em{\scshape i\kern-0.25em b}\kern-0.8em\TeX}}}

\usepackage{multirow} 

\usepackage[ruled,vlined]{algorithm2e}
\SetAlFnt{\small}

\usepackage{algpseudocode}

\usepackage{graphicx}
\usepackage{array}
\usepackage{amsmath}

\newcommand{\argmin}{\emph{argmin}}

\begin{document}

\title{Multi-Objective Fairness in Team Assembly}

\author{
Rodrigo Borges \and 
Otto Sahlgren \and 
Sami Koivunen \and 
Kostas Stefanidis \and 
Thomas Olsson \and 
Arto Laitinen
}

\institute{
Tampere University, Finland \\
 \email{\{rodrigo.borges,otto.sahlgren,sami.koivunen\}@tuni.fi}
 \email{\{konstantinos.stefanidis,thomas.olsson,arto.laitinen\}@tuni.fi}
}

\maketitle

\begin{abstract}
Team assembly is a problem that demands trade-offs between multiple fairness criteria and computational optimization. We focus on four criteria: (i) fair distribution of workloads within the team, (ii) fair distribution of skills and expertise regarding project requirements, (iii) fair distribution of protected classes in the team, and (iv) fair distribution of the team cost among protected classes. For this problem, we propose a two-stage algorithmic solution. First, a multi-objective optimization procedure is executed and the Pareto candidates that satisfy the project requirements are selected. Second, N random groups are formed containing combinations of these candidates, and a second round of multi-objective optimization is executed, but this time for selecting the groups that optimize the team-assembly criteria. 
\end{abstract}


\section{Introduction} 
Given a set of optimization criteria and constraints, team assembly algorithms are designed to select, from a pool of candidates who each have a set of skills, a set of individuals that jointly fulfils the requirements of a predefined project. Decision-makers have to establish a clear understanding of project requirements and teams' envisioned tasks to translate them into computationally tractable formal requirements, and also choose an appropriate way of assigning candidates into teams~\cite{DBLP:journals/pacmhci/HarrisGDC19}. 
Team assembly also requires attention to ethical and legal questions, especially when conducted in high-stakes domains, such as formal education or professional work contexts. 
Unfair bias, whether social or technical in origin~\cite{10.1145/3433949}, is a particularly salient concern in team assembly since it can lead to unfair treatment of candidates in the selection process and disadvantage members of protected groups (such as gender or ethnic groups).

Existing work has developed methods for improving team assembly algorithms in different respects, such as reducing the cost of team assembly~\cite{DBLP:conf/www/BarnaboFLS19}, distributing workloads more equitably among candidates~\cite{10.1145/2187836.2187950} 
and improving the representation of different groups in the resulting teams~\cite{DBLP:conf/www/BarnaboFLS19}. Whereas a large body of work is devoted to developing methods for identifying and mitigating wrongful bias and unfairness in rankings and recommendations~\cite{DBLP:journals/vldbj/pitoura21}, research addressing these issues in computational team assembly contexts remains scarce. Most existing approaches are designed for incremental solutions where teams are formed by selecting one candidate after the other, and a single distributive desideratum is optimized. Our work addresses two problems. First, decision-makers often have multiple objectives that need to be balanced simultaneously \cite{Lee2021-LEEAFI}. Second, incremental approaches to team-assembly can be undesirable in certain cases, such as when choosing one candidate at time $t_{1}$ closes off the possibility to choose another more suitable candidate at a later time $t_{2}$. 

To address these issues, we formulate team assembly as a multi-objective optimization task where fairness-aware team assembly involves considering several competing objectives. We describe our framework and illustrate its benefits by employing four criteria for fairness-aware team-assembly. 
Ideally, a team assembly algorithm would compare every possible team-composition in light of the criteria and choose the one that minimizes a target objective. However, this approach can be expensive especially when the candidate pool is large. To address this issue, we propose a two-step team assembly procedure: First, a multi-objective optimization procedure is executed and the Pareto candidates that satisfy the project requirements are selected. Second, N random groups are formed containing combinations of these candidates, and a second round of multi-objective optimization is executed, but this time for selecting the groups that optimize the team assembly criteria. The choice between teams is determined according to a combination of all fairness criteria.
This algorithm is not as cheap as selecting the best candidates incrementally, and not as expensive as testing all possible groups that can be formed. Instead, the proposed algorithm filters the best candidates among the ones that fulfill the project requirements, and forms several random groups containing these candidates. When selecting a group that is already formed one can directly access the fairness metrics, and it is easier for the algorithm to minimize a given criterion or a set of criteria.



\section{Related Work} 
A team can be defined as a set of two or more individuals interacting in a dynamic and interdependent manner towards a common goal \cite{DBLP:journals/pacmhci/Gomez-ZaraDC20}. 
Depending on the context, teams may be required to complete tasks ranging from group-learning (e.g., writing group essays) to product development and innovation processes (e.g., generating ideas~\cite{DBLP:journals/cscw/KoivunenO021}). 
Various techniques and computational solutions are available for team assembly, ranging from software systems that enable users to control the selection process to systems that optimize the process using machine learning (see \cite{DBLP:journals/pacmhci/Gomez-ZaraDC20}, \cite{DBLP:journals/pacmhci/HarrisGDC19}). 

\textit{Team assembly.} Research on computational team-assembly spans diverse application areas and different computational approaches. For example, \cite{DBLP:conf/kdd/LappasLT09} proposes a team recommender that groups individuals within a social network based on pre-defined skill requirements. 
Given a task, a pool of individuals with different skills, and a social network which captures whether individuals are compatible with one another, the goal is to define a set of users to perform the task, where the users both meet the skill requirements of the task and work effectively together as a team, using a communication cost indicator to measure effectiveness. 
\cite{DBLP:conf/msr/MintoM07} presents an approach to form and recommend emergent teams based on how software artifacts are changed by developers. In \cite{DBLP:conf/eurospi/YilmazAO15}, teams are formed based on the personality of the team members using a classifier to predict the performance of the constructed teams. \cite{DBLP:journals/corr/abs-1302-6580} frames team assembly as a group recommendation task, where a team is formed from users with specific constraints, and then items are recommended to that team.


\textit{Bias and fairness in team-assembly.} Research on software fairness has developed various metrics for identifying and addressing unfair bias. 
For example, Statistical Parity~\cite{10.1145/2090236.2090255} requires that the protected groups (e.g., gender groups) experience equal selection rates, whereas Equalized Odds~\cite{10.5555/3157382.3157469} requires that those groups experience equal error rates. Numerous techniques can be applied to mitigate bias in data, algorithms, or the output~\cite{DBLP:journals/vldbj/pitoura21}. Fairness has been discussed less in computational team-assembly contexts, though there are some notable exceptions. For example, \cite{DBLP:conf/ai/BulmerFGH20} formulates fair team assembly as a fair allocation task: students should be assigned to projects so that there are balanced workloads and tasks in the resulting teams. 
Another example is \cite{DBLP:conf/www/BarnaboFLS19}, which examines fair team-formation in an online labour marketplace. Here, each user possesses a set of skills and all users are divided into disjoint groups subject to a single fairness constraint: the formed team should have the same number of users belonging to each protected class, and the team should have all the skills needed to complete a given task. 
Lastly, \cite{DBLP:conf/webi/MachadoS19} presents the most similar setting to our work by exploring a problem where teams have multidisciplinary requirements and the incremental selection of team members is based on the match of their skills and the requirements. 

\section{Motivation}
 We propose a multi-objective approach to fairness-aware team-assembly in a subset selection setting where a single team is formed with a one-shot (non-incremental) process given a pool of candidates and a set of project requirements. This approach addresses two limitations or gaps in existing work: the focus on incremental team-assembly tasks and the use of a single fairness objective. 

\textit{Team Assembly as One-Shot Subset Selection.} We address a gap in the research literature by considering a one-shot subset selection task where a team is formed by choosing an optimal set of individuals from a larger set of candidates, where project-to-team fit is evaluated by considering project requirements and candidates' skills. This is because (1) subset selection has received comparably less attention in research on software fairness (see, however, \cite{10.1145/3514094.3534160}) and (2) existing work on fairness-aware team assembly has largely focused on an incremental approach to selecting candidates, which can undesirable or suboptimal in certain cases since the overall composition of the team can be known only after selecting all candidates. 
For example, to reduce the overall cost of the team, the incremental approach would intuitively start with selecting the cheapest candidate that satisfies one of the project requirements even though another candidate with a slightly higher cost would satisfy two other criteria with a higher score. 
 
\textit{Multi-Objective Fairness in Team Assembly.} 
Using a single fairness objective (or constraint) can result in narrow evaluations of the team-compositions and preclude identification of trade-offs between objectives~\cite{Lee2021-LEEAFI}. Our multi-objective approach to fair team assembly recognizes that team assembly procedures often distribute multiple distinct goods and opportunities simultaneously. E.g., fair team-assembly involves fair distribution of access to teams, as well as fair allocation of tasks and responsibilities between team-members, taking into account also the overall utility of the outcomes of the procedure. We specify four objectives for fair team assembly: (a) \textit{Fair Representation}: The distribution of protected attributes within a team should be as equitable as possible. 
(b) \textit{Fair Workload Distribution}: The distribution of tasks within a team should be as equal as possible. 
(c) \textit{Fair Expertise Distribution}: The distribution of skills within a team should be as balanced as possible.  
(d) \textit{Fair Cost Distribution}: The distribution of the total cost within a team should be as fair as possible considering candidates' protected attributes. 
In Table~\ref{tab:team_example}, Member 1 presents a substantially lower cost compared to others, and Req 2 sums a substantially lower cost if compared to other requirements. The members associated with class 1 concentrate a higher cost, compared to the single member associated with class 0. 
\begin{table}[t!]
 \centering
  \caption{Example of a team T formed according to project requirements P. The values in the table indicate the cost associated with each user/skill combination.  
  }
  \label{tab:team_example}
  \begin{tabular}{|c|c|c|c|c|c|} \hline
    \textbf{Members} & \textbf{Class} &\textbf{Req 1} & \textbf{Req 2} & \textbf{Req 3} & \textbf{Req 4} \\ \hline
    Member 1 & 0 & 0.000 & 0.035 & 0.000 & 0.000 \\               \hline
    Member 2 & 1 & 0.100 & 0.000 & 0.000 & 0.022 \\               \hline
    Member 3 & 1 & 0.000 & 0.000 & 0.090 & 0.081 \\               \hline
\end{tabular}
\end{table}

\section{Problem Formulation}\label{sec:prob}

Let $S =\{s_1, \dots, s_m \}$ be a set of skills, $A$ be a binary sensitive attribute that can assume values $A_0$ or $A_1$, $U = \{u_1, \dots, u_k\}$ be a set of individuals, i.e., the candidate pool, and $P$ be a set of requirements for a project, i.e., a subset of the skill set ($P \subset S$). 
An individual $u \in U$ is represented as a combination of a cost profile ($u^S$) containing the hiring cost associated with their skills, and a value ($u^A$) associated with a sensitive attribute. The cost profile is obtained through a function $\theta$ that returns the cost of a certain skill, 
for example, $u^S = (\theta(s_1,u), \theta(s_2,u), \dots, \theta(s_m,u))$ represents user $u$ according to their cost in skills $\{s_1, s_2, \dots, s_m\}$. We assume a user has a certain skill as long as the cost associated with that skill is greater than 0. 


Any set of more than 2 and less than $|U|$ individuals is considered a team $T$. And the number of project requirements that are fulfilled by a team is referred to as $coverage$. 
The aim of the team-assembly method is to select among all teams that cover the project requirements, the team that minimizes five objectives: team cost, workload uneven distribution, expertise uneven distribution, representation parity, and cost difference. These objectives are described next. 

\noindent\textbf{Team Cost.} The total cost of hiring a team for a project is defined as:
\begin{equation}\label{eq:team_cost}
Cost(T,P) = \sum_{u \in T} \sum_{j=1}^{|S \cap P|} \theta (s_j, u). 
\end{equation}

It can be described as the summation of the cost associated with each team member's skills that match the project requirements. Notably, one candidate can contribute to more than one task in the project when their skills coincide with more than one project requirement.

\noindent\textbf{Workload Uneven Distribution.} It is calculated as the standard deviation of the cost associated with each member of the team: 
\begin{equation}\label{eq:work_load}
Workload(T,P) = \sqrt{\frac{1}{|T|} \sum_{u \in T} \Big(\sum_{j=1}^{|S\cap P|} \theta (s_j, u) - \frac{Cost(T,P)}{|T|}\Big)^2}.
\end{equation}

A team in which the total cost is well distributed among members (low variance) is considered fair, and a team in which the cost is concentrated among a few members (high variance) is considered unfair.

\noindent\textbf{Expertise Uneven Distribution.}  In addition to distributing the costs fairly among candidates, the costs should also be fairly distributed among project requirements. The unevenness of expertise distribution is calculated as: 
\begin{equation}\label{eq:expertise}
Expertise(T,P) = \sqrt{\frac{1}{|P|} \sum_{j=1}^{|S\cap P|} \Big(\sum_{u \in T} \theta (s_j, u) - \frac{Cost(T,P)}{|P|}\Big)^2}.
\end{equation}

Similar to workload distribution, this objective measures the standard deviation of the cost associated with each project requirement. A fair team is expected to distribute their cost among requirements as even as possible, thus resulting in a low standard deviation value. 

\noindent\textbf{Representation Parity.} This measures the difference between the occurrences of $A_0$ and $A_1$ within a team as potential values for a sensitive attribute $A$. The objective is calculated as:
\begin{equation}\label{eq:rep_par}
Representation(T,A) = \frac{\sqrt{(|f(T,A_0)| - |f(T,A_1)|)^2}}{|T|},
\end{equation}
where function $f(T,A_0)$ returns a set containing the members of $T$ associated with sensitive attributes $A_0$, as well as in the case of attribute $A_1$. A low Representation Parity indicates a fair distribution of attribute $A$ whereas a high value indicates a majority of members associated with one of the classes, $A_0$ or $A_1$.  

\noindent\textbf{Cost Difference.} This measures the difference between the cost allocated to two categories, $A_0$ and $A_1$, within a team. The total cost of team members associated with a certain sensitive attribute, named Cost Attribute (CA), is calculated as: 
\begin{equation}
CA(T,P,A_0) = \sum_{u \in f(T,A_0)}\sum_{j=1}^{|S \cap P|} \theta (s_j, u),
\end{equation}
in the case of attribute $A_0$. And the Cost Difference objective is calculated as:
\begin{equation}\label{eq:cost_diff}
CostDiff(T,A,P) =\frac{\sqrt{(CA(T,P,A_0) - CA(T,P,A_1))^2}}{Cost(T,P)}.
\end{equation}

As mentioned before, our goal is to select a team $T$ that fulfils the project $P$ requirements and that minimizes the multi-objective condition: 
\begin{equation}
\begin{split}
\argmin_T ( Cost(T,P), Workload(T,P), Expertise(T,P), \\ Representation(T,A), 
CostDiff(T,A)).
\end{split}
\end{equation}

\subsection{Multi-Objective Fairness in Team Assembly} \label{sec:mobj} 
We propose a method designed for assembling teams with multiple fairness constraints. The method assumes a pool of candidates from which the team will be selected, a project and a sensitive attribute associated with each of the candidates. It formulates fairness-aware team-assembly as a multi-objective optimization problem performed in two stages. First, project requirements are considered as objectives, and the best candidates are selected for the next phase. Second, multiple teams are formed with these candidates and fairness constraints are calculated for each of the teams. This time the fairness constraints are assumed as objectives, and the team that minimizes these constraints while fulfilling the project requirements is considered the fairer one (Algorithm~\ref{alg:mobj}). 

\begin{algorithm}[t!]

  \caption{Algorithm (in pseudo-Python) for multi-objective fairness-aware team assembly}
  \label{alg:mobj}
  \SetAlgoLined
  \KwData{A pool of candidates (U), project requirements ($P$), team size ($M$), number of random teams (N), sensitive attribute ($A$)}
  \KwResult{the selected team (T) }
  
\texttt{\\}
 $U_p$ = FilterCandidatesWithProjectSkills($U$, $P$)\;
 paretoCandidates = MultiObjective($U_p$, min)\;  
\texttt{\\}
teams = FormNRandomTeams(paretoCandidates, N, M)
   
\texttt{\\}
candTeams = [ ]\;
 \For{team in teams}{

        {\If{$CalcTeamCoverage(team) == |P|$}{
       cost = CalcTeamCost(team, P)\;
       workload = CalcTeamWorkload(team, P)\;
       expertise = CalcTeamExpertise(team, P)\;
       representation = CalcTeamRepresentation(team, A)\;
       costDifference = CalcTeamCostDifference(team, A)\;

       \texttt{\\}
       candTeams.append([team, cost, workload, expertise, representation, costDifference])
    }
    }}

\texttt{\\}

 paretoTeams = MultiObjective(candTeams, min)\;
 T = argmax(paretoTeams)
 
 \texttt{\\}
\end{algorithm}

Given a candidate pool ($U$) and set of project requirements ($P$),
the first action is to filter candidates with at least one skill required by the project.  
($FilterCandidatesWithProjectSkills$() in Algorithm~\ref{alg:mobj}). 
The filtering process removes all users for which $|u \cap P| = 0$, and the remaining candidates are referred to as $U_p$. 
The candidates in $U_p$ are then submitted to a multi-objective optimization step with the aim of selecting the best candidates for this specific project according to the Pareto dominance concept. According to this concept, a candidate dominates another if they perform better in at least one of the project requirements. A candidate is considered non-dominated if they are not dominated by any other candidate in the population, and the set of all non-dominated candidates compose the \emph{Pareto candidates} subset.

At this point, the $paretoCandidates$ subset contains the non-dominated candidates considered as the most suitable for the given project, but our notion of a fair team can only be assessed when having a formed team. The next step is to form a reasonable amount of teams containing a fixed number of Pareto candidates selected randomly. The number of random teams ($N$) as well as the size of these teams ($M$) are provided as parameters for the method. 
Once the teams are formed, it is possible to calculate their coverage, as well as their fairness objectives with Equations~1, 2, 3, 4 and 6. 
The teams that fulfil the project requirements are filtered out 
and considered in the following step. $Cost$, $Workload$, $Expertise$, $Representation$ and $Cost Difference$ are then calculated, and each team end up being represented according to the values calculated for its objectives. 
A second round of multi-objective optimization is executed. This time, teams are being compared instead of candidates, with the same understanding as in from the first case. Given two arbitrary teams, two are the possibilities: (i) one team dominates the other if it has at least one objective with a lower value than the other team, or (ii) the two teams do not dominate each other. The non-dominated teams are referred to as \emph{Pareto teams}. 
Finally, all objectives measured for Pareto teams are summed up as an indicator of an \emph{overall unfairness}, and the method selects the team $T$ with the lower unfairness value.

\section{Experiments}

\textit{\textbf{Dataset.}} We evaluated the proposed method in a dataset obtained from the $freelancer$\footnote{\url{https://www.freelancer.com/}} website, in which candidates register themselves to be hired as freelance workers. The dataset was provided by the authors of~\cite{DBLP:conf/www/BarnaboFLS19}, and it contains 1,211 candidates who self-declared their costs and their expertise in 175 skills. In the information available in the dataset, users are associated with skills in a binary fashion (having or not having expertise), but no information is provided about the cost of each skill separately. We decided that the cost declared by the users is the same for every skill in which they have the expertise, meaning that if user $u$ declared a cost $c$ and they are hired for a project in which they will contribute with two skills, then the total cost associated with this users is $2 \times c$.

The original dataset associates each user with a cost value, and it also associates the same user with a set of skills. We decided that the price paid for hiring one user is the result of multiplying their cost and the number of skills matching the project requirement. That is to say, if user $u$ costs $c$ and they are hired to work on a project whose requirements match with one of their skills, then they will be paid $c$, but the project requirement matches with two of their skills, then they will be paid $2\times c$.

In~\cite{DBLP:conf/www/BarnaboFLS19}, the authors attributes a hypothetical binary sensitive attribute to each candidate, and generates several versions of the same dataset, associating candidates with this attribute in different proportions. For example, if we consider this sensitive attribute as gender, we could imagine a candidate poll with 70\% men, and 30\% of women. The idea is to simulate biased candidate pools and evaluate the impact of those in the teams formed by team-assembly methods. We decided to use the dataset in which members are equally represented in the candidate pool (50/50), and the dataset in which members are more unevenly represented (10/90). 
The dataset contains also the requirements for  600 projects. No information to order the projects in any way is provided, and so we consider projects independently.

\noindent\textit{\textbf{Previous Approaches.}} We applied two other team-assembly methods to the same task for the sake of comparison. The first method, named \emph{Incremental}, was inspired in~\cite{DBLP:conf/webi/MachadoS19} and selects the most suitable candidates incrementally until the project requirements are fulfilled. The second method, named \emph{Fair Allocation}, was inspired in~\cite{DBLP:conf/www/BarnaboFLS19} and operates in a similar fashion, but this time considering users associated with a sensitive attribute and forcing as much as possible that the formed team has a fair distribution of this attribute among its members.

\section{Results}\label{sec:res}

Multi-Objective, Incremental and Fair Allocation were evaluated for forming a team for each of the 600 projects in the freelancer dataset, and the average value obtained for each of the objectives presented in Section~\ref{sec:prob} was calculated. The results are reported in Table~\ref{tab:results}, separately for the two datasets containing different proportions of the sensitive attributes, 50/50 and 10/90. 
The results of Multi-Objective are presented according to five different optimization objectives, named configurations: Top-Cost, Top-Workload, Top-Expertise, Top-Representation and Top-Cost Difference, along with a Random selection variation. 
On average, the first round of multi-objective optimization reduced the number of candidates by 77\%, meaning that the candidates dominated others with compatible skills representing 23\% of the total. In the second round of multi-objective optimization, the teams were reduced by 98\% on average. The teams that dominate the others represent only 2\%. This reflects how much the teams formed randomly can be internally equivalent or redundant.   

\begin{table}[t]
\caption{Cost, Workload, Expertise, Representation, Cost Difference, and number of formed teams for Incremental, Fair Allocation and Multi-Objective methods. Multi-Objective can optimize different criteria, and its results are presented according to seven objectives: Random, Top-Cost, Top-Workload, Top-Expertise, Top-Representation, Top-Cost Difference and Top-Sum. 
The best results for each objective are in boldface, and the second-best results are in underlining.}
  \label{tab:results}
  \tiny 
  \begin{tabular}{|c|c|p{1.5cm}|c|c|c|c|c|c|} \hline

    \textbf{Classes} & \multicolumn{2}{c|}{\textbf{Algorithm}} & \textbf{Cost} & \textbf{Workload} & \textbf{Expertise} & \textbf{Representation} & \textbf{Cost} &\textbf{Teams} \\

    \textbf{Dist.} & \multicolumn{2}{c|}{} & & & & & \textbf{Difference} & \\
\hline

\multirow{9}{*}{50/50} & \multicolumn{2}{c|}{Incremental} & \textbf{23.311 (15.548)} & 0.035 (0.034) & \underline{0.025 (0.026)} & 0.480 (0.378) & 0.610 (0.345) & 486 \\
\cline{2-9}
 & \multicolumn{2}{c|}{Fair Allocation} & 29.201 (20.453) & 0.042 (0.040) & 0.032 (0.032) & 0.238 (0.268) & 0.447 (0.286) & 486 \\
\cline{2-9}
 & \multirow{7}{*}{\rotatebox[origin=c]{90}{Multi-Obj.}} 
  & Random & 40.862 (23.391) & 0.045 (0.045) & 0.035 (0.036) & 0.343 (0.339) & 0.419 (0.351) & 506 \\
\cline{3-9}
 & & Top-Cost & \underline{26.099 (16.876)} & 0.035 (0.038) & 0.026 (0.027) & 0.450 (0.360) & 0.595 (0.333) & 506 \\
 & & Top-Workload & 42.294 (27.074) & \textbf{0.018} (0.029) & 0.032 (0.036) & 0.434 (0.367) & 0.471 (0.357) & 506 \\
 & & Top-Expertise & 37.943 (23.646) & 0.039 (0.039) & \textbf{0.011 (0.020)} & 0.445 (0.363) & 0.525 (0.359) & 506 \\
 & & Top-Repres. &  28.503 (17.531) & 0.036 (0.038) & 0.028 (0.027) & \textbf{0.143 (0.170)} & 0.393 (0.251) & 505 \\
  & & Top-Cost Diff. & 41.458 (25.980) & 0.040 (0.040) & 0.037 (0.036) & 0.188 (0.197) & \textbf{0.091 (0.196)} & 506 \\
 \cline{3-9}
& & Top-Sum. & 39.147 (22.969) & \underline{0.032 (0.035)} & 0.028 (0.029) & \underline{0.144 (0.174)} & \underline{0.107 (0.206)} & 506 \\
\hline
\hline
\multirow{9}{*}{10/90} & \multicolumn{2}{c|}{Incremental} & \textbf{23.423 (15.693)} & \underline{0.035 (0.034)} & \underline{0.025 (0.026)} & 0.758 (0.341) & 0.875 (0.239) & 486 \\
\cline{2-9}
 & \multicolumn{2}{c|}{Fair Allocation} & 31.212 (20.802) & 0.046 (0.040) & 0.034 (0.030) & \textbf{0.413 (0.370)} & 0.600 (0.328) & 484 \\
\cline{2-9}
 & \multirow{7}{*}{\rotatebox[origin=c]{90}{Multi-Obj.}} 
  & Random &  40.939 (25.097) & 0.044 (0.047) & 0.035 (0.038) & 0.722 (0.354) & 0.748 (0.350) & 506 \\
\cline{3-9}
 & & Top-Cost &  \underline{26.159 (17.047)} & \underline{0.035 (0.038)} & 0.026 (0.027) & 0.729 (0.347) & 0.843 (0.253) & 506 \\
 & & Top-Workload &  42.525 (27.463) & \textbf{0.018 (0.029)} & 0.033 (0.036) & 0.828 (0.297) & 0.836 (0.285) & 505 \\
 & & Top-Expertise &  38.140 (23.855) & 0.039 (0.040) & \textbf{0.011 (0.020)} & 0.795 (0.327) & 0.842 (0.298) & 506 \\
 & & Top-Repres. & 30.231 (19.608) & 0.039 (0.041) & 0.030 (0.032) & 0.435 (0.390) & 0.594 (0.353) & 506 \\
 & & Top-Cost Diff. & 34.961 (22.256) &  0.040 (0.039) & 0.032 (0.031) & 0.470 (0.377) & \textbf{0.459 (0.439)} & 506 \\
 \cline{3-9}
 & & Top-Sum. & 37.661 (23.700) & \underline{0.035 (0.038)} & 0.027 (0.030) & \underline{0.432 (0.389)} & \underline{0.460 (0.436)} & 505 \\
\hline

\end{tabular}
\end{table}

In general, the Incremental method was the most efficient in minimizing the total cost and the size of the formed teams. The Multi-Objective method was able to assemble a slightly higher number of teams than other methods, probably because of forming teams in one-shot instead of incrementally. When configured to minimize a specific objective, the Multi-Objective method was efficient in selecting the teams, except when the objective was the Cost, and when the objective was the Representation and the dataset contained an uneven distribution of classes (10/90). In the former case, Incremental was the most efficient method, and in the latter case, the Fair Allocation performed better.

In the context of Multi-Objective fairness, it is preferable that a team presents a good balance of objectives instead of an extremely low value for one objective despite the others. E.g., when configured to optimize the Expertise objective (Top-Expertise) in the uneven dataset (10/90), Multi-Objective selected, on average, teams with a fairly high Representation (0.795) and Cost Difference (0.842) values. The Top-Sum configuration, on the other hand, selected teams with the second-best average values (highlighted with underline in Table~\ref{tab:results}) for three out of five objectives, for both datasets. The two objectives in which the configuration did not perform well were Cost and Expertise, which leads us to the next step of analyzing potential conflicts between objectives in the results.

\section{Conclusions}

In this paper, we argued in favour of a wider notion of fairness in the context of team assembly by framing the task of forming teams as a multi-objective optimization procedure. We have also proposed an algorithm for assembling teams with multiple fairness constraints that assembled teams in a one-shot fashion, as opposed to incremental methods proposed previously in the literature. Our method is flexible enough that it can be applied to situations when one single objective needs to be minimized (or maximized), as well as in situations when all objectives need to be optimized jointly. 

\bibliographystyle{splncs04.bst}
\bibliography{sample-base}


\end{document}